# Origin of the abnormal diffusion of transition metal in rutile

Linggang Zhu,[1, 2] Graeme J. Ackland,[3] Qing-Miao Hu,[4] Jian Zhou,[1] and Zhimei Sun[1, 2,*]

[1]*School of Materials Science and Engineering, Beihang University, Beijing 100191, China*
[2]*Center for Integrated Computational Materials Engineering, International Research Institute for Multidisciplinary Science, Beihang University, Beijing 100191, China*
[3]*School of Physics and Astronomy, University of Edinburgh, Edinburgh EH9 3JZ, UK*
[4]*Shenyang National Laboratory for Materials Science, Institute of Metal Research, Chinese Academy of Sciences, Shenyang 110016, China*
*corresponding author: zmsun@buaa.edu.cn*

## Abstract

Diffusion of dopant in rutile is the fundamental process that determines the performance of many devices in which rutile is used. The diffusion behavior is known to be highly sample-dependent, but the reasons for this are less well understood. Here, rutile is studied by using first-principles calculations, in order to unravel the microscopic origins of the diverse diffusion behaviors for different doping elements. Anomalous diffusion behavior in the open channel along [001] direction is found: larger atoms include Sc and Zr have lower energy barrier for diffusion via interstitial mechanism, apparently contradicting their known slow diffusion rate. To resolve this, we present an alternate model for the overall diffusion rate of the large-size dopants in rutile, showing that parallel to the [001] channel, it is limited by the formation of the interstitial states, whereas in the direction perpendicular to [001], it proceeds via a kick-out mechanism. By contrast, Co and Ni, prefer to stay in the interstitial site of rutile, and have conventional diffusion with a very small migration barrier in the [001] channel. This leads to highly anisotropic and fast diffusion. The diffusion mechanisms found in the present study can explain the diffusion data measured by experiments, and these findings provide novel understanding for the classic diffusion topic.

## Introduction

Rutile titanium dioxide is a wide bandgap semiconductor which has many applications in photocatalytic/optoelectronic devices [1,2], resistance random access memory (RRAM), etc [3-5]. It is also important as the corrosion-resistant layer on Ti alloys [6,7]. Transition metal doping is an effective method to control the bandgap and so extend and improve the application of rutile. The thermodynamic and kinetic behavior of the dopant has significant effects on the overall properties of the materials. Rutile is an example of a highly anisotropic material so understanding its properties is an important topic for basic research as well as technological application.

Typically, rutile is nonstochiometric ($TiO_{2-x}$), and the predominant intrinsic point defects are interstitial Ti ions and oxygen vacancies [8,9]. By comparing the diffusivity at different oxygen pressure, it has been concluded that oxygen and Ti migrate via the vacancy and interstitialcy mechanism, respectively. It is well known that stoichiometry has a large effect on dopant diffusion properties, but whether this is due to the charge on the migrating defects, trapping by intrinsic defect or some other mechanism is poorly understood.

In rutile, each Ti is surrounded by 6 oxygen atoms forming a slightly distorted octahedron. These TiO6 octahedra share edges and corners, and the Ti ions in the center of the octahedra lie in rows parallel to $c$ axis. When rutile is viewed along $c$ axis, the "open channels" surrounded by these octahedra can be seen. Thus the structure of rutile is highly anisotropic: the open channel along $c$ axis ([001] direction) may provide a fast diffusion path for interstitial defects. There are no such channels perpendicular to the $c$ axis, and self-diffusion in that direction proceeds via the interstitialcy mechanism (also called the kick-out mechanism), which involves a sequence of collisions and replacements of lattice titanium ions by Ti interstitials. In the case of dopants, the equivalent kick-out mechanism involves simultaneous movement of one titanium ion and one dopant. These two distinct mechanisms mean that anisotropic diffusion can be expected.

Extensive experimental studies on the diffusion of various elements in rutile have been performed and reported in the literature [8,10-14]. Sasaki et al [10] reported a comprehensive study of the diffusion of Sc, Cr, Mn, Fe, Co, Ni and Zr, using the radioactive-tracer sectioning technique. They found that the migration of Co and Ni is extremely anisotropic, with the diffusion along $c$ axis much faster than that perpendicular to $c$ axis. However, the other cations show weaker anisotropy, as does Ti self-diffusion. By considering the diffusion coefficient at various oxygen pressure and temperature, Sasaki et al [10] also investigated the diffusion mechanism of the cations: for Sc, Zr, and Cr, the diffusion coefficients are strongly coupled with the concentration of interstitial Ti, which suggests the interstitialcy mechanism; Nb has identical diffusion behavior to the self diffusion of Ti [12] also suggesting that the interstitialcy mechanism dominates. By contrast, for Ni and Co, direct migration of the interstitial in open channels was believed to account for the rapid diffusion along the $c$ axis.

Although some deductions can be made based on the analysis of the experimental data, critical parts are still missing for a complete picture of dopant diffusion. Firstly, it is clear from the atomic structure of rutile that interstitialcy and interstitial mechanisms account for diffusion perpendicular and parallel to $c$ axis, respectively. Thus for elements without apparent anisotropic diffusion behavior, these two mechanisms must coexist. This raises a curious problem: in the interstitialcy mechanism, the dopants kick-out a Ti and transform from a substitutional defect to an interstitial defect, it is then the Ti atom, not the dopant, which is able to migrate further. By contrast, interstitial migration along the channel allows repeated jumps of the same atom. The possibility that dopants can be trapped on substitutional sites means that the two dominant diffusion mechanisms in rutile are correlated. Their coupling or competition cannot be easily resolved by the experiments, so theoretical calculations are required disentangle the effects by studying the interstitialcy and interstitial mechanisms separately.

Theoretical studies of diffusion of intrinsic defects in rutile are also abundant in the literature [15-18], but a comprehensive study on the diffusion of different dopants is still missing. In the present work, by combination of first-principles calculations and transition state theory, the microscopic mechanisms for the anisotropic diffusion of the transition metals in rutile, including Sc, Ti, V, Ni, Co, Nb and Zr, are revealed.

## Calculation Details

All the calculations are performed by using VASP (Vienna Ab initio Simulation Package) [19,20]. The generalized gradient approximation (GGA) parameterized by Perdew and Wang (PW91) [21] is used to describe the electronic exchange-correlation potential. The plane-wave cutoff energy is set as 400 eV and the ions are described using PAW pseudopotentials. We use a supercell of 240 atoms with periodic boundary conditions along [001], [110], and $[1\bar{1}0]$ directions: this has previously proven large enough to obtain a converged diffusion barrier for interstitial defects [15]. The supercell is kept fixed while the atoms are free to relax during the optimization. The k-point mesh is 2×2×2, generating 4 irreducible k-points. All the calculations are spin-polarized. The interstitial and interstitialcy mechanisms are shown in Fig.1. For the evaluation of the diffusion path and barrier, the climbing image Nudged Elastic Band (NEB) method is used [22]. To check on the importance of localization of 3$d$ electrons, some comparison calculations using GGA+U and hybrid functional HSE06 [23] are performed to validate the results based on GGA.

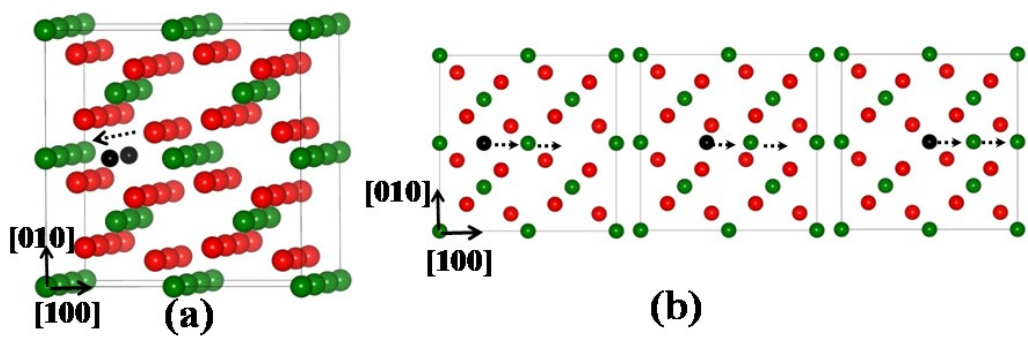

Fig. 1 Dominant diffusion mechanisms in rutile (a) interstitial mechanism in the direction parallel to $c$ axis (b) interstitialcy (kick-out) mechanism in the direction perpendicular to $c$ axis. Red and green balls represent the O and Ti atoms, while black ball is the diffusing atom.

## Results and Discussions
### A: Diffusion along $c$ axis

Diffusion of the dopant along *c* axis is along the open channel in *c* direction via the interstitial mechanism. The two high-symmetry interstitial sites in the open channel are shown in Fig. 2. The octahedral site is normally the stable site for interstitial atoms, while the 4-fold coordinated site is typically the transition state for hopping between adjacent octahedral sites.

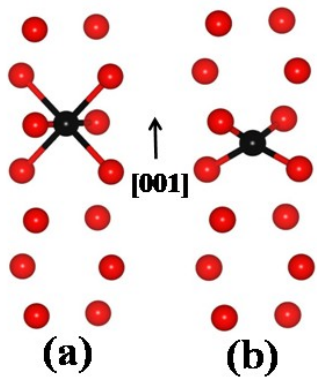

Fig. 2 Interstitial states in the channel formed by oxygen atoms along [001]. (a) Octahedral site (b) 4-fold coordinated site. Red balls represent the O atoms, while black ball is the diffusive atom.

The migration energy barriers calculated using GGA are given in Fig. 3, together with the atomic radii of the dopants. The energy profiles for the diffusions are shown in Fig. S1 in the Supplemental Material [24]. For the self-diffusion of Ti, we calculated an energy barrier of 0.61 eV, in fair agreement with previous calculation (0.70 eV) [15]. The striking and surprising feature of Fig. 3 is that the energy barrier does not increase with the atomic radius. Some of the larger atoms have low barriers, implying faster diffusion, which contradicts the experiment. For example, for group IVB elements, 3*d* element Ti has a higher energy barrier than the much larger sized 4*d* element Zr. The same trend is found for group VB elements V and Nb. Comparing 3*d* elements, the larger Ti and Sc atoms have lower migration barriers than V, while even the smallest atoms, Co and Ni, have comparable barriers to Sc.

Unlike other dopants, Ni prefers the 4-fold coordinated site in the channel with the octahedral position being the transition site. This feature might be due to the much smaller size of Ni; a similar situation has been found for hydrogen in iron, which prefers the tetrahedral site while large atoms such as C and P occupy the octahedral site [25,26].

Electronic structure calculations of transition metal oxides may suffer from strong electron correlation effects. To check the effects of the *d*-electron localization, we repeated the calculations for Ti and Ni diffusion using a Hubbard U correction as implemented in Liechtenstein's method [27]. The U parameters are taken from the literature [28], where they are shown to reproduce the thermodynamic property of different oxides values and are applied to the *d* electrons of Ti (U=2.0eV, J=1.0eV), and Ni (U=3.4 eV; J =1.0 eV). The resultant energy barriers are 0.81 eV and 0.24 eV for Ti and Ni, respectively. Although the Hubbard U term leads to a higher energy barrier, the discrepancy is not significant enough to change the trend in Fig. 3, and the discussions below are all based on the GGA calculations.

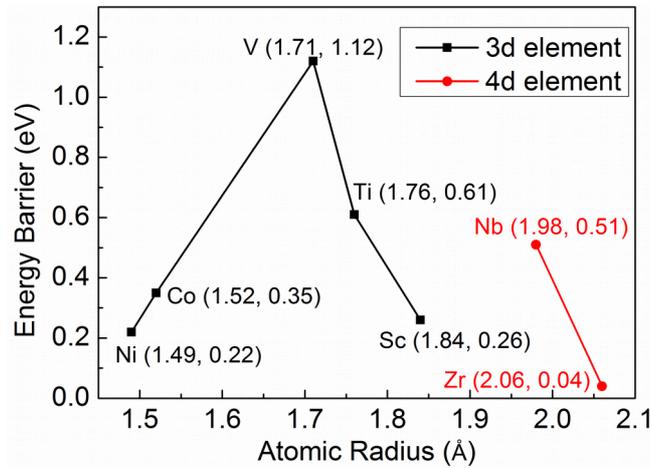

Fig. 3 Calculated energy barriers using GGA functional for the diffusion of the dopants versus their radii [29]. The element is labeled with the value of its atomic radius and corresponding energy barrier in the parenthesis.

In the following part, we will discuss the physical origin of the abnormal diffusion along *c* axis, i.e., why the larger atoms including Sc, Nb and Zr have low diffusion barriers along *c* direction, yet have low diffusion rates. In metallic Ti [30] and in the semiconductor CdTe [31] as well as other semiconductors [32], unexpected diffusion behaviors have been ascribed to features such as indirect diffusion paths between initial and final states, specific orbital coupling between the host and dopant atom, etc. In rutile we found that the diffusion paths for interstitial dopants are all linear and parallel to [001] direction and all the dopants are transition metals with well-localised *d*-electron configurations. A more likely explanation is that both the octahedral and the 4-fold interstitial states have high but similar energy, with the substitutional site being more stable for large dopants, even when intrinsic Ti interstitials are present. If both interstitial sites are metastable, only a small fraction of the dopants will be in those states. Consequently, the low diffusion rate is due to the low number of diffusers.

To investigate further, we compare the electronic density of states (DOS) of the initial and transition structures (Fig. 4). The conventional formal charges in rutile are $Ti^{4+}$ and $O^{2-}$, which results in an empty conduction *d*-band on the cation and an occupied *sp*-type valence band on the oxygen. When an interstitial dopant atom is added, it introduces extra *d*-electrons to the system. These either remain localised on the dopant as gap states, or become delocalised and occupy the bottom of the conduction band. The red curves in Fig. 4 show the new states induced by the doping.

For Ni, there are several gap states, and in the octahedral site (transition state as found in NEB calculations), the conduction band is partly occupied, while in 4-fold coordinated sites (right panel) an additional localized energy level near -1 eV is occupied. This agrees with the finding that octahedral site is less stable and act as the transition state for the diffusion of Ni.

For V and Ti octahedral sites, a gap state (-0.5 eV for V and -0.2 eV for Ti) is found close to conduction band. This state moves to a higher level when the dopants diffuse into 4-fold coordinated transition sites. For the larger Sc and Zr, the DOS for the initial and transition states are very similar, consistent with the lower barriers, and the extra electrons always stay in the conduction band for both of the initial and transition structures, resulting in a small energy difference between the two states, i.e., lower energy barrier. In fact, such a low migration barrier for an oversized defect, has also been observed in Ti [30], Fe [33], and in AlN single crystals in recent experimental study [34], indicating that this counterintuitive diffusion phenomenon is not that rare in nature.

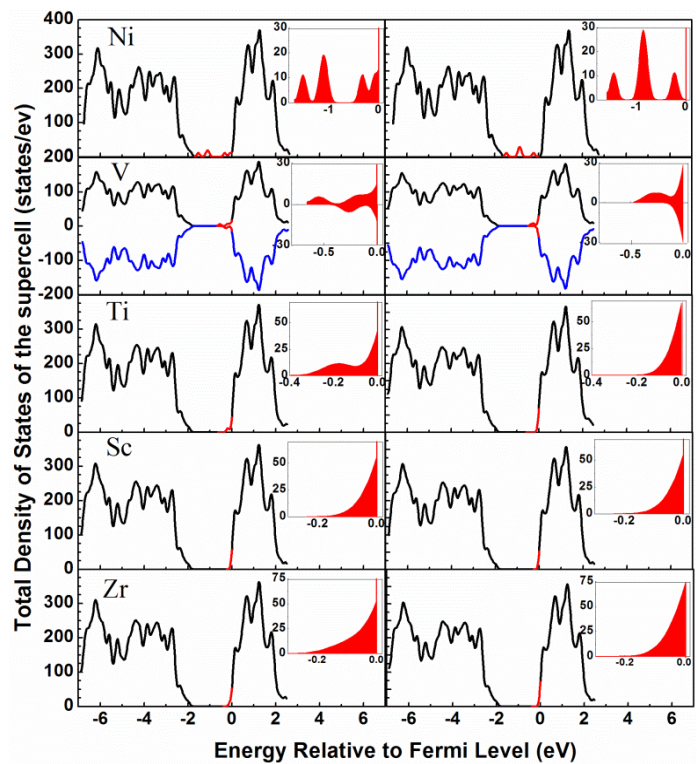

Fig. 4 Total Density of States for rutile with different dopant. The left panels correspond to the dopant in the octahedral sites (Fig. 2(a)), while the right panel is for the dopant in the 4-fold coordinated sites (Fig.2(b)). States which are not present in pure $TiO_2$ are highlighted using red line and shown magnified in the inset.

The gap states are related to the charge state of the migrating ion. Within DFT, this can be studied using projected density of states (pDOS) shown in Fig. 5. pDOS is not uniquely defined because of charge "belonging" to the oxygen $sp^3$ shell overlapping into the region associated with the projection functions. However, it shows how the migrating and interstitial ions are charged. The pDOS of the dopants shows that there are essentially no *s*-electrons remaining on any of the cations, which rule out any explanation in terms of *s-d* coupling [32]. Sc and Zr have no *d*-electrons remaining in the gap, and, therefore are 3+, and 4+ ions, respectively. By integrating the pDOS, we find that V still has one *d*-electron in the gap state, so migrates as a 4+ ion, while Ni retains 6 *d*-electrons in the gap states, once again behaving as a 4+ ion. There is no significant difference between the charge state in the interstitial and migrating states.

We investigated the effect on migration of varying the chemical potential by repeating some calculations with supercells containing fewer electrons and a neutralizing jellium background. In most cases, where the jellium is replacing delocalised conduction band states, the migration barrier is unchanged, such as Sc. The one notable exception is Ti. As seen in Fig 5, Ti has a gap state in the octahedral site some 0.3 eV below the conduction band. The fourfold site has no such gap state. So to allow migration in substoichiometric rutile one electron must be excited into the conduction band (Ti3+ in the initial state becomes Ti4+ atop the barrier). For the low-chemical potential calculation (i.e. using a supercell with fewer electrons) the gap state is unoccupied, the excitation is unnecessary, and the migration barrier is lowered to 0.30 eV. Thus we can expect the experimental self-diffusion of Ti in rutile to depend very sensitively on sample composition. Diffusion profiles for neutral/charged Sc and Ti can be seen in Fig. S2 in the Supplemental Material [24]. More discussions about the charged defect can be found in the APPENDIX.

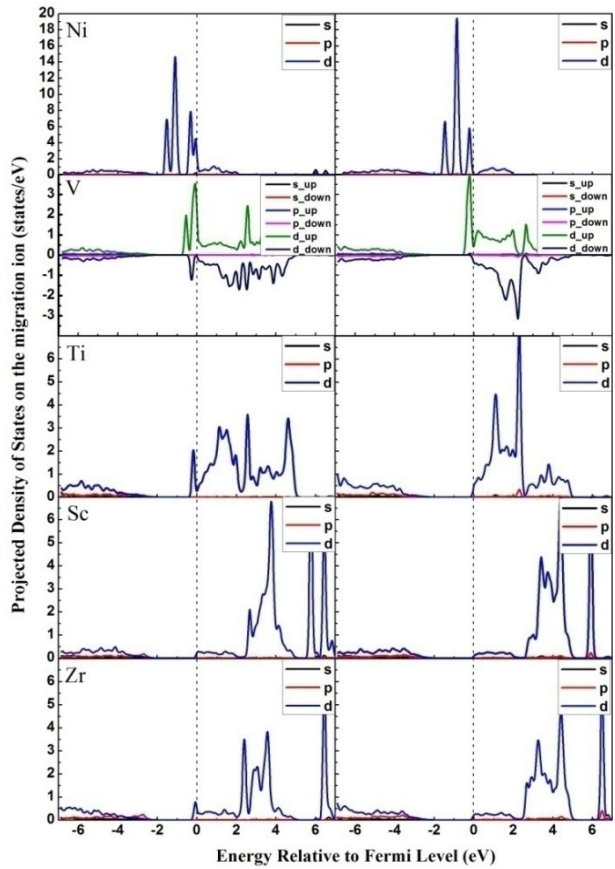

Fig. 5 DOS projected onto the migrating atom in the doped rutile. The left panels correspond to the situation when the dopant stays in the octahedral sites (Fig. 2(a)), while the right panel is the DOS of the structures with dopant in the 4-fold coordinated transition sites (Fig.2(b)). The vertical dash-line indicates the position of the Fermi Level.

Normally, it is expected that atoms with smaller migration barriers will diffuse faster. For example, according to Fig. 3, Zr might be expected to diffuse faster than Ni and Co along *c* direction. However, this is opposite to the experimental observation. The important feature which we have not yet considered is the formation energy of the "defect" that is needed for the diffusion process, e.g. in vacancy mediated diffusion the vacancy formation energy determines how many vacancies are available to diffuse.

The formation energy of the "defect" (interstitials) is sometimes ignored because the concentrations of the interstitial dopants is temperature-independent, being determined by the sample stoichiometry. In rutile, the substitutional site is generally the most stable site for any cation, however, if there are excess cations, then some of them must be in interstitial sites. The interstitial atom can be either an intrinsic Ti defect, or the dopant ion. We define the *kick-out process* as the formation of an interstitial dopant from a substitutional dopant plus an intrinsic Ti intersitial. The concentration of interstitial dopants depends heavily on the energy the kick-out and will be studied in detail in the following sections.

## B: Interstitial dopant *vs* substitutional dopant

As we mentioned above, the pre-existing interstitial Ti in non-stochoimetric rutile enables the diffusion process perpendicular to the c-axis via the kick-out mechanism. Through this mechanism, the dopant switches between the substitutional site, and an interstitial site in the open channel. The kick-out energy, $E_{diff}$, is defined as the energy difference between the system with an interstitial dopant and the one with an interstitial-Ti together with a substitutional dopant. A negative $E_{diff}$ indicates that the dopant prefers the interstitial site to the substitutional site. Fig. 6 shows that $E_{diff}$ depends on the atomic radius, with dopants smaller than Ti stable in the interstitial site.

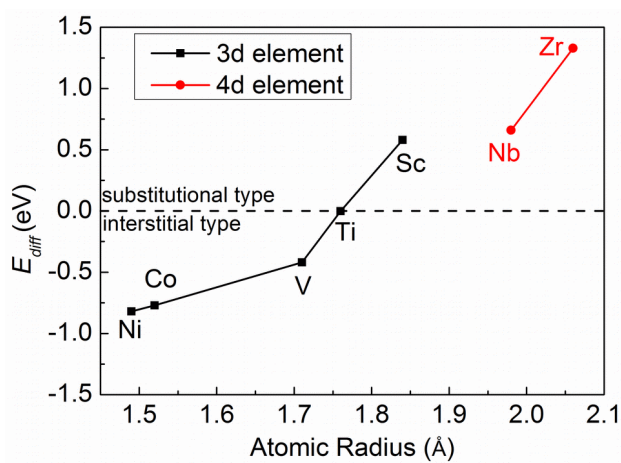

Fig. 6 Relationship between atomic radius of the dopant and $E_{diff}$, defined as the energy difference between structures where the dopant is interstitial and those in which the dopant is substitutional and accompanied by a Ti interstitial [see also Fig.1(b)]. Positive $E_{diff}$ indicates that the dopant tends to replace the lattice Ti, creating a Ti interstitial, while a negative value means that the dopant prefers to stay in the interstitial site.

The trend in Fig. 6 is further validated using hybrid functional HSE06 based on a 72-atom supercell with Γ-point sampling. Firstly, for the rutile with the dopants in the interstitial site in the *c* direction, we find that GGA and HSE06 calculations lead to distinct magnetisms (Table S1 in the Supplemental Material [24]): GGA results indicate that the doped systems are non-magnetic except when the dopant is Co; while HSE06 results show that the doped rutile are all magnetic. However, for the occupation preference of the dopants, HSE06 can reproduce the trend obtained by GGA (Fig. S3 in the Supplemental Material [24]). This means that when the energy difference between two structures is what matters, such as the kick-out energy or the diffusion barrier, GGA is reliable compared to HSE06 calculations, or DFT+U calculations as shown in the study of oxygen vacancy diffusion in rutile [18].

For diffusion parallel to the *c* direction, the number of interstitial dopants is critical and the kick-out energy determines the fraction of the dopants in the interstitial site. Sc, Nb and Zr favour substitutional sites, therefore, they spend most of their time trapped in the substitutional site, but migrate quickly once in the interstitial site. Creation of dopant interstitials is thermally activated, so for these elements the effective migration barrier along the *c*-axis is the sum of the kick-out energy and the diffusion barrier energy.

## C: Interstitial dopant generated by kick-out mechanism: diffusion perpendicular to *c* axis

We take Ti, V, Sc, Nb and Zr as examples to study the diffusion perpendicular to the *c* axis. The energy profile for the migration via the kick-out process is shown in Fig. 7. It can be seen that the energy curve is not smooth, with a local minimum corresponding to the structure in the middle of Fig. 1(b). For Ti, the migration barrier for diffusion perpendicular to *c* axis is 0.60 eV, slightly lower than that along *c* direction (0.61 eV), in good agreement with the experimental data [35] and other calculations [15]. The migration of the V atom from the stable interstitial site to the substitutional transition site has a high energy barrier of 1.42 eV, while the reverse migration (trapping) has a barrier of 0.94 eV. Formation of Sc, Nb and Zr interstitials from the stable substitutional site requires overcoming energy barriers of 0.93 eV, 1.16 eV and 1.52 eV, respectively, whereas the energy barriers are respectively only 0.35 eV, 0.5eV and 0.19 eV for retrapping. The effect of this high energy barrier for the larger atom to enter the interstitial site on the diffusion will be analyzed in the next section.

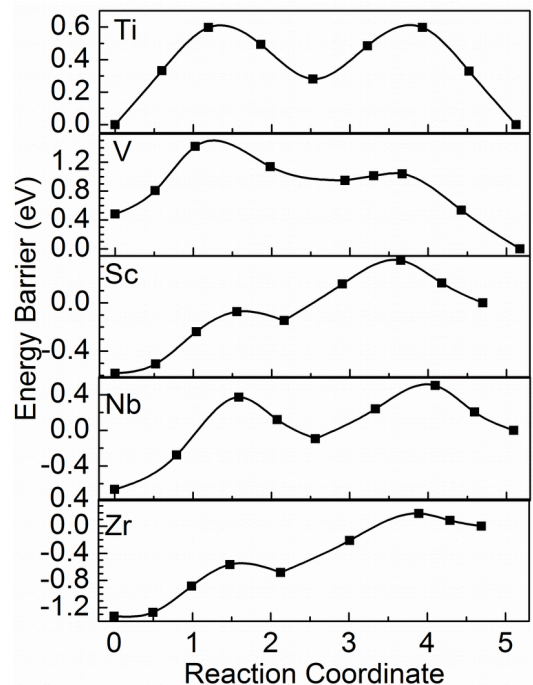

Fig. 7 Diffusion profile of dopant in the direction normal to *c* axis via kick-out mechanism. The squares are the calculated data, and the smooth curve is obtained by spline function fitting. The diffusion pathway can be found in Fig. 1(b): during the calculations the diffusion paths normal to *c* axis are divided into two parts with the separation at the structure showing in the middle figure of Fig. 1(b) (corresponding to the local minima in the energy curves), and three intermediate images are used for each separated diffusion path.

**D: Diffusion anisotropy in comparison with experiment**

In the previous sections we showed that dopants diffuse along *c* axis via interstitial mechanism whereas the kick-out mechanism governs the diffusion perpendicular to the *c* axis. In this part, we compare our results with available experimental observations [10], focusing on the anisotropy of the diffusion. The details of the experimental data can be seen in Table S2 in the Supplemental Material [24].

(1) For the self-diffusion of Ti, the calculated energy barriers are 0.61 and 0.60 eV for the diffusion in the directions parallel and perpendicular to the *c* axis, via interstitial and kick-out mechanisms, respectively. Thus the diffusion of Ti in rutile shows very weak anisotropy, in good agreement with the experimental measurement.

(2) For Co and Ni, the migration energy barriers along *c* axis via interstitial mechanisms are 0.35 and 0.22 eV, respectively. Moreover, as shown in Fig. 6, they preferentially occupy the interstitial sites of non-stoichiometric rutile, which also means that Co and Ni are unlikely to kick out the lattice Ti and diffuse perpendicular to the *c* axis. Therefore, diffusion of Co and Ni along *c* axis will be much faster and strong diffusion anisotropy is expected, in good agreement with the experimental results [10].

(3) Although V also prefers the interstitial site of rutile, the calculated migration barrier in the direction parallel to *c* axis is as large as 1.12 eV, whereas the energy barrier for the diffusion of V atom perpendicular to *c* axis is 1.42 eV. The relatively small difference between these migration energy barriers for V means that anisotropy of the diffusion of V is not as significant as that of Co and Ni.

(4) Sc, Nb and Zr, show lower migration barriers in the *c* direction. However, data in Fig. 6 indicate that interstitial sites in the channel along *c* direction are unstable for these atoms. Diffusion of these atoms along *c* direction requires the kick-out process to create an interstitial, which can then migrate quickly. In Fig. 7, we show that considerable energy is needed for the formation of these interstitial dopants via kick-out mechanism. Therefore, for diffusion either perpendicular or parallel to the *c* axis, the kick-out mechanism is the rate limiting step, indicating that preexisting interstitial Ti is also essential. This agrees well with the experimental conclusion that the diffusion of Sc and Zr are slow and coupled to the concentration of interstitial Ti [10].

# Conclusions

In the present work, the diffusion of the transition metal atoms in rutile is studied by using first-principles calculations in combination with transition state theory. The diffusions along the directions parallel and perpendicular *c* axis via interstitial and interstitialcy mechanisms, respectively, are investigated separately. The migration energy barrier for the diffusion along the *c* direction exhibits an abnormal trend, i.e., larger atoms experience lower energy barrier. However, it is shown that for these large atoms, the overall diffusion along *c* axis is inhibited by the high energy of the formation process of the interstitial dopants which occurs by the same kick-out process as the diffusion perpendicular to *c* axis. This provides a comprehensive theoretical picture for the strong diffusion anisotropy of Co and Ni in rutile (fast along the c axis and slow perpendicular to it), and the weak diffusion anisotropy of V, Ti, Sc, Nb and Zr. The strong anisotropy causes the bulk diffusion to be highly sensitive to the texture of the sample. More importantly, we find that, for Sc, Nb and Zr, the kick-out process affects diffusion both parallel and perpendicular to the *c* axis, and as a result, their diffusions are coupled with the concentration of interstitial Ti in rutile (i.e the non-stoichiometry). Our calculations explain the experimental observations of both rate and anisotropy of the diffusion of transition metals in rutile by introducing the kick-out mechanism. It is expected that this type of mechanism operates in many oxide materials, leading to similar sensitivity to the stoichiometry.

## Acknowledgement

This work is partially supported by National Natural Science Foundation of China (51401009, 61274005). Prof. Sun thanks the support from National Natural Science Foundation for Distinguished Young Scientists of China (51225205). Dr. Linggang Zhu acknowledges the ERC training network MAMINA. Prof. Ackland acknowledges the financial support from EPSRC via grant K01465X and the Royal Society for a Wolfson award. QMH thanks to the financial support from China MoST under grant No. 2016YFB0701301.

## APPENDIX: Charged defect in DFT

In principle, density functional theory does not require the definition of electronic wavefunctions or orbitals. Furthermore, the use of periodic boundary conditions and Bloch's theorem do away with the concept of localized electrons. One must therefore be careful in defining what is meant by a charged defect.

Supercell DFT calculation can readily be performed on cells where the number of electrons is not equal to the total ionic charge. In most codes, the number of electrons is defined and the divergent Coulomb term at G=0 is simply ignored: this is the so-called "jellium" approach. If the number of electrons is fixed, then the system adjusts its chemical potential to minimize the free energy. Given only the density, the "ionic charge" can be defined by integrating the electronic density within a region surrounding the ion: defining this region is somewhat arbitrary, common choices being spheres of fixed radius, Voronoi polyhedra, or the Bader method.

The issue of boundary conditions is critical to this discussion. The boundary condition for the electrons is set by choosing the number of the electrons in the cell as a whole: in the absence of localized orbitals it is not possible, even in principle, to set up a calculation with a fixed charge on any particular ion. In our present calculations, the purpose of the periodic boundary condition is to represent the large region of perfect rutile surrounding the defect. This standard method is, nevertheless imperfect in several ways. The defect itself is repeated periodically, and so can interact with its own image, a problem which can be eliminated by going to sufficiently large supercells. More problematic for semiconductors is that the boundary condition for electrons should be defined by the external chemical potential, not by the number of electrons in the supercell.

In the Kohn-Sham approach, wavefunctions are reintroduced into DFT. Strictly, these wavefunctions represent non-interacting pseudoelectrons, but they are still constrained by the system symmetry and are typically regarded as being meaningful, allowing band structures and densities of states to be calculated. These wavefunctions can be used to define the ionic charge by projecting them onto atom-centered orbitals. Again, there is some arbitrariness in choosing these orbitals by well-establish methods exist, e.g. the Mulliken charge method.

Thirdly, one can use "change in polarization when an ion is displaced" to define an ionic charge: this is the so-called Born Effective Charge, which is actually a second rank tensor.

Fourthly, one can add a dopant ion to stoichimetric $TiO_2$, choose the overall number of electrons, and assume that the net charge in the supercell is the charge on the ion. There is nothing to guarantee that this will be the case.

Finally, for our calculations, it is normal to find one or more wavefunctions localized on the dopant with an energy level in or close to the band gap. Such a wavefunction is absent in pure $TiO_2$. The occupation of this level can be used to define the ionic charge.

Thus there are no fewer than five different ways to define ionic charge and, unfortunately, they seldom give the same answer.

In Fig. 4 and Fig. 5, we use the density of states calculation to determine the charge state of the atom. In each calculation we set the chemical potential such that the supercell is neutral. For stoichimetric $TiO_2$ the sp3 oxygen band lies below the Fermi level, while the *sd* titanium states are above it – this is unambiguously a calculation in which all ions are charged. When dopants are added, the additional electrons are either spontaneously localized on the dopant ion, or delocalized into the conduction band, as clearly shown in Fig. 4. For neutral supercells, the Fermi energy shifts into the conduction band, which ensures that the charge state of the cation is insensitive to the chemical potential. The extra electrons remain within the supercell, which means that our diffusion calculations are done with some 0.02 conduction electrons per atom, corresponding to a high external electric potential. For lower chemical potential (i.e., charged supercells), the gap states in Ni, V, and Ti may be unoccupied. This leads to a high sensitivity of the barriers to chemical potential, and therefore to sample preparation. For example, Fig. 5 shows that Ti in the octahedral site has a gap state, which is absent in the 4-fold site. When the chemical potential is high enough that the state is occupied the barrier is 0.61 eV. This could be regarded as being because Ti3+ must be oxidized to Ti4+. By contrast, with a lower chemical potential (calculation with a 4+ charged supercell) this gap state is never occupied, the charge state is always Ti4+ and the barrier is reduced to 0.30 eV. In contrast for Sc, Fig. 5 indicates that no gap states exist for the octahedral site as well as the 4-fold site, meaning that regardless of supercell charge we are dealing with a Sc3+ ion. A calculation based using a 3+ charged supercell leads to a barrier of 0.24 eV, equivalent to that of the neutral defect (0.26 eV). Diffusion profiles for neutral/charged Sc and Ti can be seen in Fig. S2 in the Supplemental Material [24].